\documentstyle[psfig]{mn0}

\def\simg{\mathrel{\rlap{\raise 0.511ex \hbox{$>$}}{\lower 0.511ex \hbox{$\sim$}}}}
\def\siml{\mathrel{\rlap{\raise 0.511ex \hbox{$<$}}{\lower 0.511ex \hbox{$\sim$}}}}
\def\d{\rm d} \def\E{{\cal{E}}}

\begin{document}

\title[Swift GRB Afterglows]
   {Swift Gamma-Ray Burst Afterglows and the Forward-Shock Model}

\author[A. Panaitescu]{A. Panaitescu \\
        Space Science and Applications, MS D466, Los Alamos National Laboratory, Los Alamos, NM 87545, USA}

\maketitle

\begin{abstract}
\begin{small}
 The X-ray light-curves of the GRB afterglows monitored by Swift display one to four phases of 
power-law decay. In chronological order they are: the burst tail, the "hump", the standard decay, 
and the post jet-break decay. We compare the decay indices and spectral slopes measured during
each phase with the expectations for the forward-shock model to identify the processes which may 
be at work and to constrain some of their properties.
 The large-angle emission produced during the burst, but arriving at observer later, is consistent
with the GRB tail decay for less than half of bursts. 
 Several afterglows exhibit a slow, unbroken power-law decay from burst end until 1 day, showing 
that the forward-shock emission is, sometimes, present from the earliest afterglow observations. 
In fact, the forward-shock synchrotron emission from a very narrow jet (half-angle less than 
$1^{\rm o}$) is consistent with the decay of 75 percent of GRB tails. The forward-shock inverse-Compton 
emission from a narrow jet that does not expand sideways also accommodates the decay of 80 percent of 
GRB tails.
 The X-ray light-curve hump can be attributed to an increasing kinetic energy per solid angle of the 
forward-shock region visible to the observer. This increase could be due to the emergence of the 
emission from an outflow seen from a location outside its opening. However, the correlations 
among the hump timing, flux, and decay index expected in this model are not confirmed by observations.
Thus, the increase in the forward-shock kinetic energy is more likely caused by some incoming ejecta 
arriving at the shock during the afterglow phase.
 The jet interpretation for the burst tails and the energy injection scenario for the hump lead to 
a double-jet outflow structure consisting of a narrow GRB jet which precedes a wider afterglow 
outflow of lower kinetic energy per solid angle but higher total energy.
%(X-ray light-curves are shown for GRB 050416A, 050525A, 050717, 050721, 050724, 050730, 050802, 
%051109A, 060124, 060206, 060526).
\end{small}
\end{abstract}

\begin{keywords}
  gamma-rays: bursts - radiation mechanisms: non-thermal - shock waves
\end{keywords}

\section{Introduction}

 The X-ray, optical and radio emission following a Gamma-Ray Burst (GRB) is thought to arise in the 
interaction between the GRB ejecta and the circumburst medium, which leads to a forward shock energizing 
the ambient medium (the "external shock model" - e.g. Paczy\'nski \& Rhoads 1993, M\'esz\'aros \& Rees 1997). 
This shock accelerates electrons (through first order Fermi mechanism or electric fields associated with 
the Weibel instability) to relativistic energies and generates magnetic fields (e.g. Medvedev \& Loeb 1999). 
The afterglow emission 
is synchrotron; inverse Compton scatterings may affect the electron radiative cooling and contribute 
to the early X-ray afterglow emission. The progressive, power-law deceleration deceleration of the 
forward shock leads to a continuous softening of the afterglow synchrotron spectrum. 
As this spectrum is a combination of power-laws, $F_\nu \propto \nu^{-\beta}$ (with the {\it spectral 
slope} $\beta$ depending on the location of the observing frequency relative to the afterglow characteristic 
break frequencies), it follows that the afterglow light-curve decays as a power-law, $F_\nu \propto 
t^{-\alpha}$ (with the {\it decay index} $\alpha$ depending on $\beta$ and the evolution of the spectral 
characteristics).

 In its simplest form, the standard forward-shock model assumes a GRB outflow with a constant energy, 
a uniform kinetic energy per solid angle, and constant microphysical parameters. The possibility
of energy injection into the forward shock was proposed by Paczy\'nski (1998) and Rees \& M\'esz\'aros 
(1998). Its effect may have been observed for the first time in the rise of the optical emission of 
GRB afterglow 970508 at 1 d (Panaitescu, M\'esz\'aros \& Rees 1998). The effect of ejecta collimation 
was treated by Rhoads (1999) and may have been seen for the first time in the optical light-curve of 
GRB afterglow 990123 (Kulkarni et al 1999). Since then, about a dozen of other optical afterglows 
displayed a break at around 1 day (e.g. Zeh, Klose \& Kann 2006), which have been interpreted as evidence 
for tightly collimated outflows. A non-uniform angular distribution of ejecta kinetic energy per solid 
angle was proposed by M\'esz\'aros, Rees \& Wijers (1998) and identified by Rossi, Lazzati \& Rees (2002) 
as a possible origin for optical light-curve breaks.

\begin{figure}
\centerline{\psfig{figure=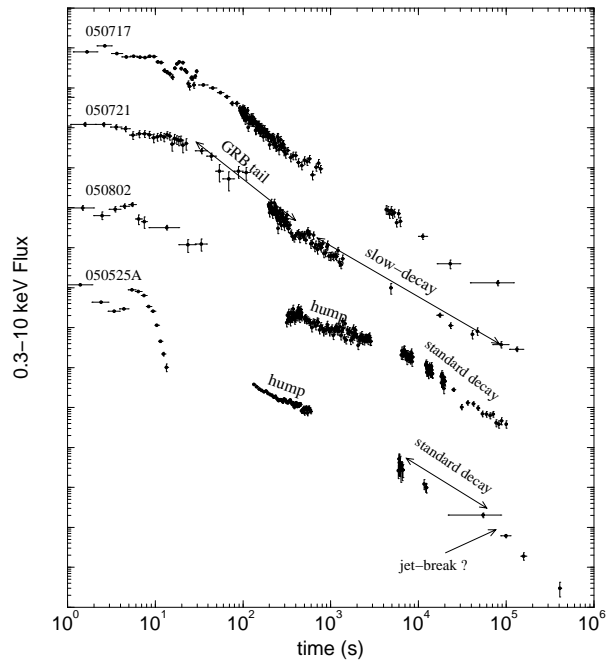,width=8cm}}
\caption{ The three types of X-ray afterglows observed by Swift: after the burst (10--30 s), 
      the light-curve displays $(i)$ a single power-law decay (top),
      $(ii)$ a steeper decay (GRB tail) followed by a break to a slow decay (second from top),
      $(iii)$ a phase of very slow decay (hump) between the GRB tail and a standard decay phase 
             (third from top).
     Several afterglows exhibit a second break to an even steeper decay (bottom).}
\label{types}
\end{figure}

 The continuous monitoring during the first day by the Swift satellite has shown that GRB X-ray afterglows
exhibit up to four decay phases (Figure \ref{types}). 10 percent of Swift afterglows exhibit a single 
power-law decay ($\alpha_x \sim 1.5$), from end of burst to about 1 day. A quarter of afterglows show a 
steeper decay ($\alpha_{x1} > 1.75$) after the burst (the "GRB tail"), followed by a break to a slower 
power-law fall-off ($0.5 < \alpha_{x2} < 1.25$) until after 1 day. About two-thirds of afterglows exhibit 
an even slower decay ($0 < \alpha_{x2} < 0.75$) after the GRB tail, followed by a steeper fall-off ($0.75 
< \alpha_{x3} < 1.75$), creating a "hump" in the X-ray light-curve at 1--10 ks. The X-ray light-curve of
several afterglows displays a second break at $\sim 1$ day, followed by a steeper decay ($1.6 < \alpha_{x4} 
< 2.4$).

 In this work, we compare the decay indices and spectral slopes of 78 X-ray afterglows with the expectations 
of the forward-shock model and discuss some of the mechanisms which may be at work during the four possible 
afterglow phases. 
 The existence of long-lived, power-law decaying light-curves is a natural prediction of the forward-shock
model (e.g. M\'esz\'aros \& Rees 1997) arising from $(i)$ the power-law deceleration of a relativistic 
blast-wave and $(ii)$ the power-law distribution with energy of particles accelerated at shocks.
 The forward-shock emission depends on the outflow dynamics (determined by the blast-wave energy \& 
collimation and medium density) and radiation emission parameters (two microphysical parameters quantifying 
the electron and magnetic field energies and the index of the power-law distribution of post-shock electrons 
with energy). It follows that the decay of the forward-shock emission depends on the evolution of the kinetic 
energy per solid angle $\E$ of the visible part of the outflow, the ambient medium stratification, and the 
possible evolution of microphysical parameters. 

 The X-ray afterglows used here were monitored by Swift from January 2005 to the end of July 2006. 
Their X-ray decay indices and spectral slopes are taken from O'Brien et al (2006) and Willingale et al (2007). 

\section{GRB Tails}
\label{tail}

\begin{figure*}
\vspace*{-10mm}
\centerline{\psfig{figure=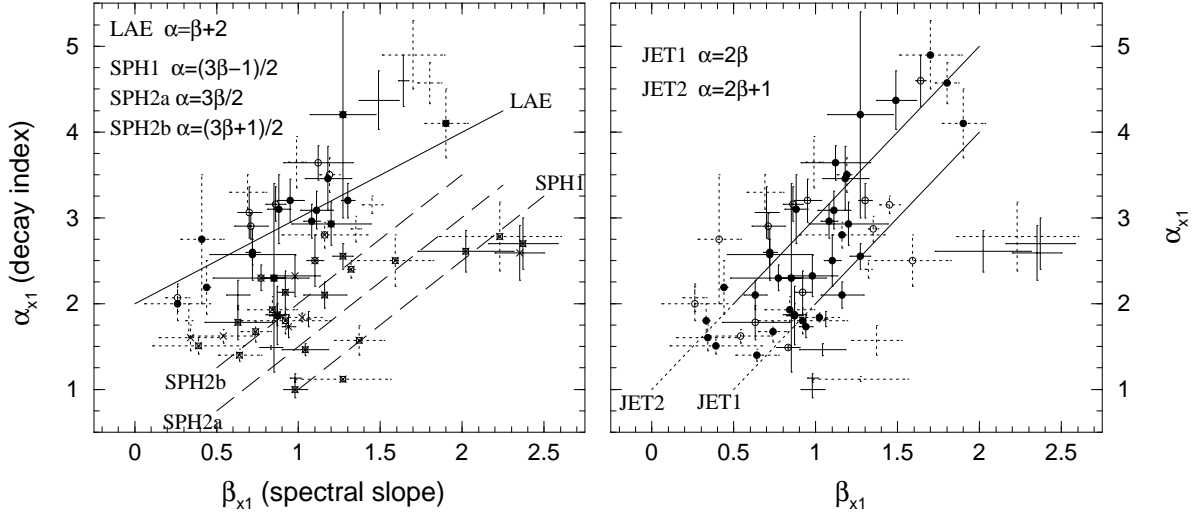,height=8cm,width=16cm}}
\caption{ Decay index $\alpha$ vs. spectral slope $\beta$ ($F_\nu \propto t^{-\alpha} \nu^{-\beta}$) 
     for the tail emission of 63 Swift GRBs (including 2 short bursts) compared with the expectations
     for the large-angle emission model (LAE) and the {\bf synchrotron} emission from a decelerating 
     forward shock.
     Label "SPH" is for a spherical outflow (in the sense that its lateral boundary is not yet visible), 
     "JET" is for a spreading jet whose boundary is visible to observer.
    Derivations of the model $\alpha-\beta$ relations can be found in M\'esz\'aros \& Rees (1997), 
     Sari, Piran \& Narayan (1998), Chevalier \& Li (1999), Rhoads (1999), Kumar \& Panaitescu (2000).
    Here and throughout this article, label "1" is for cooling frequency below X-ray, 
    "2" for cooling frequency above X-ray, "a" is for a homogeneous circumburst medium, 
     and "b" for a wind-like medium (radial stratification $n \propto r^{-2}$).
    For cooling frequency below X-ray (label "1") and synchrotron emission, the X-ray light-curve
     decay index is independent of the stratification of the ambient medium (i.e. models SPH1a and
     SPH1b yield the same decay index $\alpha$).
   Filled and empty circles indicate bursts whose decay index $\alpha_{x1}$ is consistent within 
    $1\sigma$ and at $1\sigma-2\sigma$ with the LAE (left panel) or JET (right panel) model expectations, 
    respectively. Encircled stars and stars show the bursts consistent with SPH model at the respective
    levels.
   GRB tail decays which are not consistent within $2\sigma$ with any model have no symbol.
   Solid error bars are for X-ray afterglows with a hump, dotted for those without.
  }
\label{x1}
\end{figure*}

 Figure \ref{x1} compares the decay indices and spectral slopes during the GRB tail with the expectations 
from different models. The correlation of $\alpha_{x1}$ and $\beta_{x1}$ is statistically significant 
($r = 0.46 \pm 0.05$ for 63 bursts, corresponding to a less than 0.1 percent probability of a chance 
correlation) and represents a natural consequence of any model in which, during the GRB tail, the 
spectral break frequencies of the synchrotron spectrum decrease. 

\begin{figure*}
\vspace*{-5mm}
\centerline{\psfig{figure=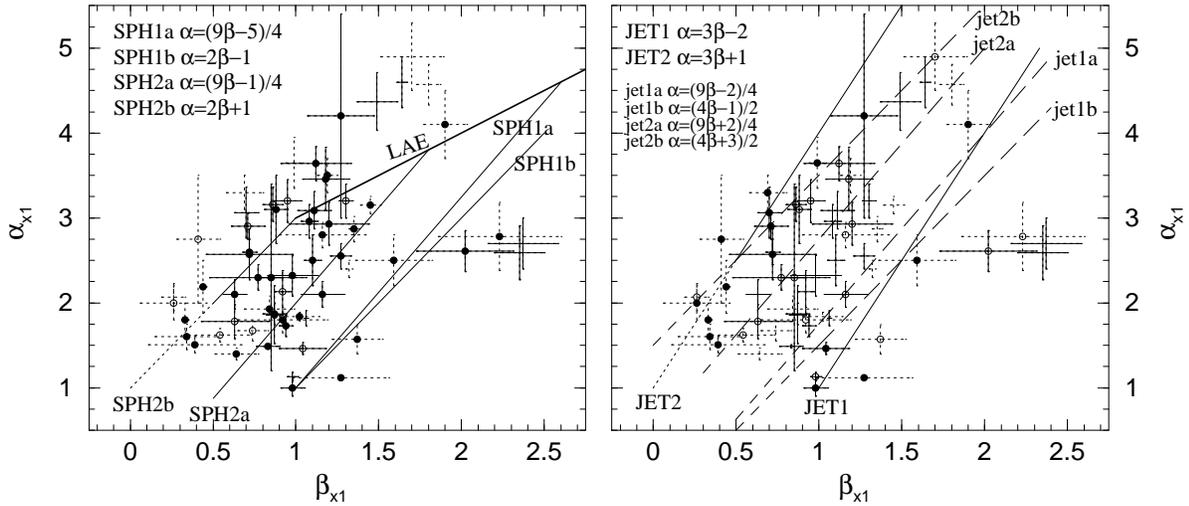,height=7.5cm,width=16cm}}
\caption{ Comparison between the decay index -- spectral slope relation expected for the 
     {\bf inverse-Compton} emission from a decelerating forward shock and observations of GRB tails.
     {\it Left panel:}  spherical outflow; the model $\alpha-\beta$ relations follow from equations
      (65) and (66) of Panaitescu \& Kumar (2000). For a spherical outflow, the steepest decay possible 
       is that of the large-angle emission, hence the "cut" set by the LAE model.
     {\it Right panel:} narrow jet; for the JET model (laterally spreading outflow), the 
       $\alpha-\beta$ relations have been derived from the equations for jet dynamics (Rhoads 1999)
       and inverse-Compton spectral characteristics (Panaitescu \& Kumar 2000, Sari \& Esin 2001). 
       Models labelled "jet" are for an outflow whose boundary is visible and which does 
       not expand laterally (conical jet). For these models, the decay index $\alpha$ follows from 
       those for a spherical outflow (left panel), corrected for a multiplying factor $\Gamma^2$ 
       accounting for that the visible source subtends a solid angle $\propto \Gamma^{-2}$ for a 
       spherical outflow and constant for a conical jet (Panaitescu, M\'esz\'aros \& Rees 1998).
       In each panel, filled and open symbols show the GRB tails consistent with the SPH or JET 
       models within $1\sigma$ and at $1\sigma-2\sigma$, respectively. This symbol coding is used
       for the rest of the article.   
  }
\label{x1ic}
\end{figure*}

 One way to discriminate the possible models for this phase is the collimation of the GRB outflow. 
If the outflow opening $\theta_0$ is larger than $\Gamma^{-1}$, the inverse of its Lorentz factor, 
then the spherical forward-shock (SPH) models shown in Figure \ref{x1} can explain only the slower
decaying GRB tails. For the rest, their steeper decay requires that the GRB emission mechanism switches 
off at the end of the burst. The steepest decay that can be obtained by a switch-off has an index 
$\alpha_{x1}=2+\beta_{x1}$ (Fenimore \& Sumner 1997) because any faster cessation will be overshined 
by the emission from the 
fluid moving at an angle $\theta$ (relative to the center--observer direction) larger than $\Gamma^{-1}$.
The above decay index of this large-angle emission (LAE model -- Kumar \& Panaitescu 2000) is due to 
the photon arrival time increasing as $\theta^2$, while the relativistic Doppler boost decreases as 
$\theta^{-2}$. The latter also induces a dependence of $\alpha_{x1}$ on $\beta_{x1}$, as photons of 
a fixed observer frequency correspond to an increasingly larger comoving frequency. 

 The LAE model is consistent at the $1\sigma$ level with 25 percent of the GRB tail decays shown in 
the left panel of Figure \ref{x1} and consistent with 40 percent of afterglows at the $2\sigma$ level, 
where consistency at the $n\sigma$ level between a model index $\alpha_{model} = a\beta_x + b$ and an 
observed $\alpha_x$ is defined by $\alpha_x-\alpha_{model}$ being within $n\sigma = 
n [\sigma(\alpha_x)^2+a^2\sigma(\beta_x)^2]^{1/2}$ of zero [$\sigma(\alpha_x)$ and $\sigma(\beta_x)$ 
are the $1\sigma$ measurement errors]. 
 It fails to accommodate the slower-decaying tails and a few faster ones. 
 Slower decays indicate that the burst emission does not cease sufficiently fast to reveal the large-angle 
emission. 

 Faster decays indicates that the large-angle emission does not exist, i.e. the outflow opening, 
$\theta_0$, is narrower than the relativistic beaming cone, $\Gamma^{-1}$, and the GRB tail decay 
reflects the intrinsic dimming of the burst emission. Taking into account that the Lorentz factor 
$\Gamma$ of a decelerating blast-wave of isotropic-equivalent kinetic energy $E = 10^{53}\, E_{53}$ 
ergs, interacting with a WR stellar wind, is $\Gamma = 60\, E_{53} [(z+1)/3.5]^{1/4} (t/100\, 
{\rm s})^{-1/4}$, the underlying condition $\theta_0 < \Gamma^{-1}$ implies very narrow jets, 
with $\theta_0 \leq 1^{\rm o}$.

 The GRB tail emission could arise from internal shocks occurring in a variable outflow (Rees \& 
M\'esz\'aros 1994) if those shocks continue to occur after the prompt emission phase. This possibility 
is supported by that existence of flares during many GRB tails, whose short timescale is inconsistent 
with a forward-shock origin (e.g. Zhang et al 2006). 
 However, it seems very unlikely that internal shocks could account for the smooth, GRB tails lasting 
from 100 s to 10-100 ks, as seen for GRBs 050717, 050826, 051006, 051021B, 051117B, and 060403
(the most spectacular case is GRB 061007 -- Schady et al 2007 -- whose afterglow power-law decay 
extends over 4 decades in time, from 100 s to $10^6$ s).

 Alternatively, the fast decay of the GRB tails could be the emission from a decelerating forward 
shock, with internal shock yielding only the flaring emission. As shown in the right panel of Figure 
\ref{x1}, the synchrotron emission from a spherical outflow decays too slow to accommodate the GRB
tails. Provided that the shock microphysical parameters are sufficiently small, the forward-shock
inverse-Compton emission may peak at $\sim 100$ keV. Its decay is faster and, as shown in the left
panel of Figure \ref{x1ic}, is consistent within $1\sigma$ with 60 percent of GRB tails (80 percent 
within $2\sigma$). Attributing the burst emission to inverse-Compton scatterings also has the
advantage that it can explain the low-energy spectra harder than the synchrotron optically-thin, 
$F_\nu \propto \nu^{1/3}$ emission (Panaitescu \& M\'esz\'aros 2000, Stern \& Poutanen 2004).

 Faster decays for the forward-shock emission can be obtained if the GRB outflow is a narrow jet
whose boundary is visible to the observer. 
 We find that decay of the synchrotron emission from such a narrow jet undergoing lateral spreading 
is consistent with 55 percent of the bursts at the $1\sigma$ level and with 75 percent at $2\sigma$ 
(right panel of Figure \ref{x1}). 
 The right panel of Figure \ref{x1ic} shows that the decay of the forward-shock, inverse-Compton 
emission from a narrow, spreading jet is consistent within $1\sigma$ with 30 percent of GRB tails 
(60 percent within $2\sigma$). A better description of the GRB tail decay is obtained if the jet 
does not spread laterally (conical jet), the consistency percentages being 45 and 80, respectively.
 These models fails to explain only the slowest GRB tails, which should be attributed to the forward-shock 
emission from outflows wider than $\Gamma^{-1}$.

\section{Slow-Decays and Humps}
\label{hump}

\begin{figure}
\centerline{\psfig{figure=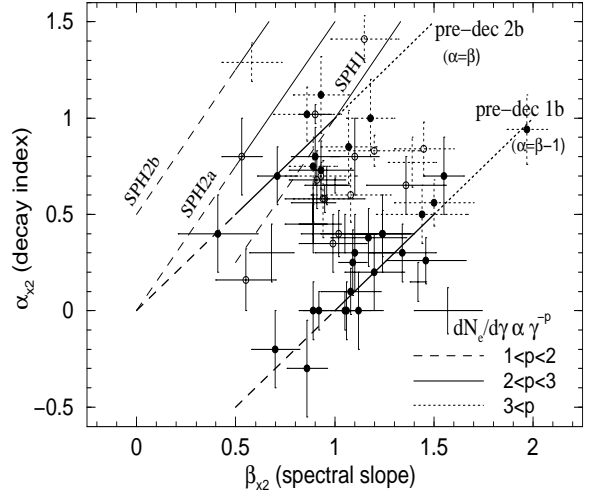,height=7cm,width=8cm}}
\caption{ Decay index vs. spectral slope during the slow-decay phase following the GRB tail, for 55 
    Swift bursts. Afterglows whose X-ray light-curves exhibit a hump are shown with solid error bars, 
    while dotted error bars indicate afterglows with a slow-decay phase whose end was not observed
    until the last Swift observation ($\sim$100 ks). The X-ray humps exhibit, on average,
    a slower decay ($\overline\alpha_{x2} = 0.37 \pm 0.30$) than the afterglows with only
    a slow-decay following the GRB tail ($\overline\alpha_{x2} = 0.86 \pm 0.25$), although the
    average spectral slopes are comparable ($\overline\beta_{x2} = 1.02 \pm 0.26$ and 
    $\overline\beta_{x2} = 1.16 \pm 0.32$, respectively).
   Two-thirds of afterglows decay slower than expected for the SPH model, the standard adiabatic 
    blast-wave model with constant microphysical parameters, whose $\alpha(\beta)$ relations are 
    given in figure \ref{x1}. 
   Thick lines labeled "pre-dec 1b" and "pre-dec 2b" are for the forward-shock emission from a 
    spherical outflow interacting with a wind-like medium, {\it before} the forward shock begins 
    to decelerate (\S\ref{predec}). 
   }
\label{x2}
\end{figure}

 We assume that the microphysical parameters pertaining to the emission from the forward-shock are 
constant. Then the X-ray hump or slow-decay phase require an increasing kinetic energy per solid angle 
$\E$ over the visible surface. This is illustrated in Figure \ref{x2}, which shows that, for an 
adiabatic forward-shock, the slowest decay (obtained for a spherical outflow) is too fast to explain 
this phase. 

 There are two reasons for a non-constant kinetic energy per solid angle in that part of the 
blast-wave which is visible to the observer: the radial and angular distribution of energy in the
GRB ejecta. In the former case, an increase of $\E$ will result before all the GRB ejecta  
undergo deceleration due to their interaction with the circumburst medium.
In the later case, the average $\E$ over the region visible to the observer ($\theta < \Gamma^{-1}$)
changes with time as the outflow is decelerated and the observer receives emission from an ever 
wider part of the outflow.

\subsection{Ejecta Deceleration and Energy Injection in the Forward Shock}
\label{enginj}

\begin{figure*}
\centerline{\psfig{figure=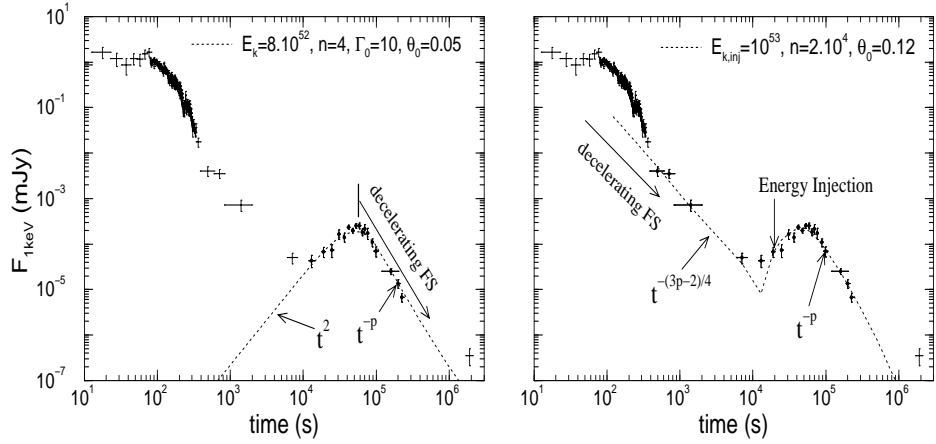,height=6cm,width=12.5cm}}
\caption{ {\it Left panel}: the onset of deceleration yields a peaked forward-shock synchrotron 
     light-curve if the circumburst medium is homogeneous, which can explain the X-ray emission 
     of the {\sl short}-GRB 050724 at 10--200 ks, provided that the jet boundary becomes visible 
     to the observer around the deceleration onset epoch (i.e. the outflow opening $\theta_0$ is 
     close to $\Gamma_0^{-1}$, 
     where $\Gamma_0$ is the ejecta pre-deceleration Lorentz factor), to explain the following
     steep ($t^{-p}$) decay afterward. The condition $\theta_0 \sim \Gamma_0^{-1}$ implies that, 
     if the burst and afterglow arise from the same outflow, then the GRB tail is not the large-angle 
     emission. 
     {\it Right panel}: the same brightening of the X-ray afterglow 050724 can be accommodated 
     by a substantial, episodic energy injection in a {\it decelerating} forward-shock. 
     The emission prior to the injection can also account for part of the GRB tail.
     A narrow jet is also required in this case by the sharp decay observed after 60 ks. 
     {\it Both panels}: legend gives the isotropic-equivalent of the ejecta kinetic energy in ergs,
     medium density in protons per ${\rm cm^{-3}}$, and jet opening in radians. } 
\label{0724}
\end{figure*}

\subsubsection{Ejecta with single Lorentz factor}
\label{predec}

 One possible way in which energy is injected into the forward shock is that where, after internal 
collisions, the relativistic ejecta move at nearly the same Lorentz factor $\Gamma_0$. The transfer 
of kinetic energy from the cold ejecta to the circumstellar medium lasts until the reverse shock crosses 
the ejecta shell. 
 
 If the comoving-frame ejecta density ($n_{ej}$) is larger than $4 \Gamma_0^2 n$ (where $n$ the 
circumburst medium density) then, during this phase, the Lorentz factor $\Gamma$ of the shocked medium
is constant ($\Gamma \siml \Gamma_0$) and the forward-shock light-curve decay is determined only by 
the increasing number of radiating electrons ($N_e \propto n r^3$, with $r$ the shock radius) and the 
decreasing magnetic field strength ($B \propto \Gamma n^{1/2}$) for a wind-like circumburst medium. 
Relating $r$ to the observer time $t$ through $r \propto \Gamma^2 t$, the spectral characteristics 
of the received synchrotron emission -- peak flux $F_p \propto N_e B \Gamma$, peak frequency $\nu_p 
\propto \gamma_p^2 B \Gamma$, and cooling frequency $\nu_c \propto \gamma_c^2 B \Gamma$, where 
$\gamma_p \propto \Gamma$ is the typical energy of the shock-accelerated electrons and $\gamma_c 
\propto \Gamma/(B^2 r)$ is the energy of the electrons whose radiative cooling time equals the dynamical 
timescale -- have the following evolutions before decelerations: $F_p = t^3$, $\nu_i = const$, 
$\nu_c \propto t^{-2}$ for a homogeneous medium and $F_p = const$, $\nu_i = t^{-1}$, $\nu_c \propto t$ 
for a wind medium ($n \propto r^{-2}$). Taking into account that, for a power-law distribution with
energy above $\gamma_p$ of the accelerated electrons ($\d N_e/\d\gamma \propto \gamma^{-p}$), the
synchrotron flux at frequency $\nu$ is $F_\nu = F_p (\nu_p/\nu)^\beta$ with $\beta = (p-1)/2$ for
$\nu_p < \nu < \nu_c$ and $F_\nu = F_p (\nu_p/\nu_c)^{\beta-1/2} (\nu_c/\nu)^\beta$ with $\beta = p/2$ 
for $\nu > \nu_p,\nu_c$, the X-ray light-curve decay index before deceleration is:
\begin{equation}
 \alpha_x\,(\nu_p < \nu < \nu_c)  = - 3 \quad , \quad  
 \alpha_x\,(\nu > \nu_p,\nu_c) = -2
\label{s0}
\end{equation}
for a homogeneous medium ($\alpha < 0$ means a rising light-curve) and 
\begin{equation}
 \alpha_x\,(\nu_p < \nu < \nu_c)  =  \beta_x  \quad , \quad   
 \alpha_x\,(\nu > \nu_p,\nu_c) =  \beta_x-1 
\label{s2}
\end{equation}
for a wind.

 If $n_{ej} < 4 \Gamma_0^2 n$ then $\Gamma = (\Gamma_0/2)^{1/2} (n_{ej}/n)^{1/4}$ (equation 5 of 
Panaitescu \& Kumar 2004). Assuming that the ejecta do not spread radially, $n_{ej} \propto r^{-2}$. 
This means that for a wind-like medium ($n \propto r^{-2}$) the ratio $n_{ej}/n$ is constant and so 
is $\Gamma$, i.e. equation (\ref{s2}) holds in this case as well. For a homogeneous medium, $\Gamma 
\propto n_{ej}^{1/4} \propto r^{-1/2}$ leads to $\Gamma \propto t^{-1/4}$, i.e. there shocked medium
is decelerated even before the reverse shock crosses the ejecta and most energy is transferred to the
forward sock. After repeating the above derivation, we find that $\alpha_x = \beta_x - 1$ for either 
location of cooling frequency, i.e. the same $\alpha_x - \beta_x$ relation as for a wind-like medium 
in the $\nu_c < \nu$ case.

 The pre-deceleration, rising X-ray light-curve resulting for a homogeneous medium (equation \ref{s0}) 
can explain the brightening of the short-GRB afterglow 050724 at 10 ks (Figure \ref{0724}), 
however long-lived brightenings are very rare. Most of the humps and all slow-decays exhibit
a falling-off emission thus, if they are attributed to pre-deceleration forward shock and
a single-$\Gamma$ ejecta, a {\it wind}-like circumburst medium is required. As illustrated 
in Figure \ref{x2}, the decay of the pre-deceleration forward-shock emission resulting for a 
wind-like medium is consistent within $1\sigma$ with 55 percent of the humps and slow-decays 
indices measured by Swift and with 80 percent of them within $2\sigma$. 

 If the ejecta shell is geometrically thick ($\Delta > r/\Gamma_0^2$) then the deceleration 
timescale (defined as the time when the reverse shock crosses the shell) for a wind-like medium 
is, in the observer frame, $t_{dec} = 0.71 (z+1) \Delta/c$ (Panaitescu \& Kumar 2004). Hence, 
in this pre-deceleration model for the X-ray light-curve hump, the source of relativistic ejecta 
would have to operate for a duration comparable to the time $t_b/(z+1)$ when the hump ends, 
i.e. for 1--10 ks.
If the engine operates for a shorter time, then the ejecta shell is thin and $t_b$ constrains 
the ejecta Lorentz factor $\Gamma_0$. For a wind-like medium, the deceleration timescale is 
$t_{dec} = 6\,(z+1) E_{53} A_*^{-1} \Gamma_{0,2}^{-4}$ s (equation 21 in Panaitescu \& Kumar 2004, 
with $t_{dec}$ twice larger to account for the arrival time of photons emitted by the fluid moving 
at angle $\theta = \Gamma_0^{-1}$ relative to the center--observer direction), using 
the $X_n = 10^{-n} X$ notation, with $X$ in cgs units. Here $E$ is the ejecta isotropic-equivalent 
kinetic energy and $A_*$ is a measure of the wind density: $n(r) = 3\times 10^{35} A_*\, r^{-2}$ 
($A_*=1$ for the wind blown by a star with a mass-loss rate of $10^{-5} M_\odot \,\rm{yr^{-1}}$
and a terminal wind velocity of $1000\; {\rm km\;s^{-1}}$). Hence, the ejecta Lorentz factor is
\begin{equation}
 \Gamma_0 = 150\, \left( \frac{E_{53}}{A_*} \right)^{1/4} \left( \frac{t_b}{z+1} \right)^{-1/4} 
\label{g0}
\end{equation}
In the case of a thick shell, the Lorentz factor of equation (\ref{g0}) would represent a lower limit 
for $\Gamma_0$.
                                                  
\begin{figure}
\centerline{\psfig{figure=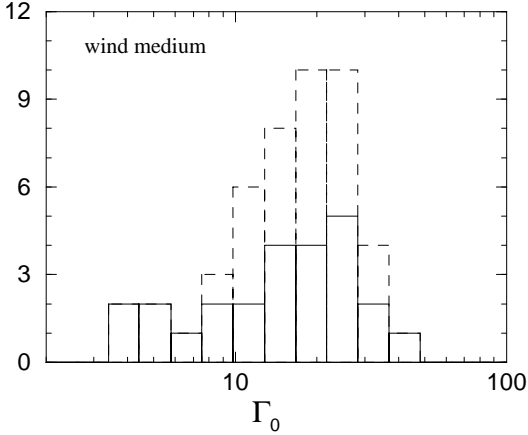,width=7cm}}
\caption{ Distribution of ejecta Lorentz factor $\Gamma_0$ for which the deceleration timescale
   is equal to the time when the X-ray light-curve hump ends, inferred from equation (\ref{g0}).
   Solid histogram is for 25 bursts with known redshift, the dashed histogram is for a set
   including 22 more bursts for which z=2.5 was assumed (an error of $\Delta z=1.5$ implies an
   error of 55 percent in $\Gamma_0$). The averages and dispersions of these two distributions are
   nearly the same: $\overline\Gamma_0 = 17 \pm 9$. }
\label{G0s2}
\end{figure}

 Assuming that the ejecta energy $E$ is comparable to the 10 keV--1 MeV burst output (which can 
be calculated from the burst fluence and redshift) and that $A_*=1$ (as for a WR star), 
we obtain the distribution of $\Gamma_0$ (for which $t_{dec} = t_b$) shown in Figure \ref{G0s2}. 
The average for 25 bursts with an X-ray hump and known redshift is $\overline\Gamma_0 = 17 \pm 10$. 
We note that an error by a factor 3 in the ejecta kinetic energy $E$ or wind parameter $A_*$
implies an error of a third in $\Gamma_0$, which is half of the dispersion of $\Gamma_0$ among
various bursts. Thus the uncertainty of $E$ and $A_*$ is unlikely to change significantly the 
distribution shown in Figure \ref{G0s2}.

\subsubsection{Ejecta with a spread in Lorentz factor}
\label{spread}

 Another variant of energy injection in the forward shock is that where, after internal interactions 
in the outflow, the ejecta do not move at a single Lorentz factor. Internal shocks order the ejecta 
Lorentz factors increasing outward, energy injection occurring after the leading edge of the outflow 
begins to decelerate and the inner shells start to catch up with it. The kinematics of this process 
is such that the arrival at the forward shock of all ejecta carrying significant energy can last much 
longer than the central engine lifetime. 
The right panel of Figure \ref{0724} illustrates how an episode of substantial energy injection 
into a decelerating forward shock can explain the brightening of GRB afterglow 050724. 

\begin{figure}
\centerline{\psfig{figure=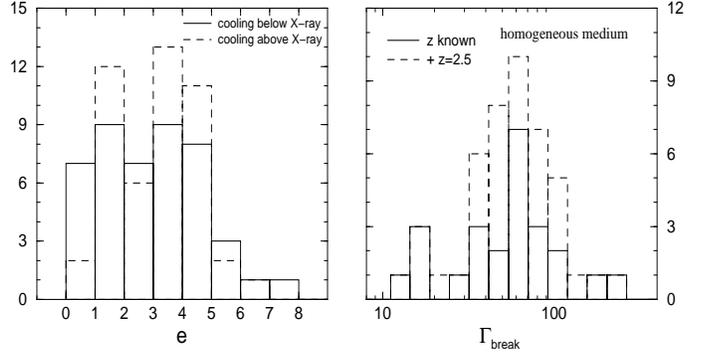,width=9cm,height=5cm}}
\caption{ {\it Left panel}: distribution of exponent $e$ of the law $E(>\Gamma_i) \propto \Gamma_i^{-e}$ 
   for energy injection in the forward shock that mitigates its deceleration and accommodates the
   slow decay of the light-curve humps of 45 X-ray afterglows. A homogeneous circumburst medium was
   assumed, for which the index $e$ is given in equation (\ref{es0}).
   Solid histogram is for $\nu_c < \nu_x$, dashed for the opposite case. 
   For either case, the parameter $e$ has a large dispersion: $\overline{e} = 2.9 \pm 1.7$ 
    for $\nu_c < \nu_x$ and $\overline{e} = 3.1 \pm 1.3$ for $\nu_x < \nu_c$. 
  {\it Right panel}: distribution of the break Lorentz factor, $\Gamma_{break} \equiv \Gamma_i(t_b)$
   (equation \ref{gbreak}),
   below which the incoming ejecta do not carry a significant energy, such that the X-ray hump 
   ends at $t_b$, when the ejecta moving at $\Gamma_i(t_b)$ arrive at the forward shock.
   Solid histogram is for 24 afterglows with measured redshift, dashed histogram is for the entire set 
   of 45 X-ray afterglows with humps, assuming $z=2.5$ for those without redshift. The break Lorentz 
   factor has a wide distribution: $\overline\Gamma_0 = 65 \pm 45$.  }
\label{eGs0}
\end{figure}

 This model for the X-ray light-curve hump is similar to that where all ejecta move at a single
Lorentz factor in that, for both, the hump lasts until all ejecta undergo deceleration, but it
differs in that the forward shock is decelerated during the hump, albeit its deceleration is
mitigated by the energy injection. One can constrain the distribution of ejecta kinetic energy with
Lorentz factor from the observed decay index and spectral slope of the X-ray hump. Assuming,
for simplicity, that the cumulative ejecta energy is a power-law in the ejecta Lorentz factor,
$E (>\Gamma_i) \propto \Gamma_i^{-e}$ ($e >0$ for a decelerating forward shock) and taking into
account that, for a short-lived engine, the forward-shock Lorentz factor is proportional to that
of the incoming ejecta ($\Gamma/\Gamma_i = [(e+2)/(e+8)]^{1/2}$ for a homogeneous medium and
$[(e+2)/(e+4)]^{1/2}$ for a wind), the condition of adiabatic dynamics for the forward shock
($\Gamma^2 n r^3 \propto E$) leads to the following dynamics: $\Gamma \propto t^{-3/(e+8)}$
for a homogeneous medium and $\Gamma \propto t^{-1/(e+4)}$ for a wind. Repeating the calculation
of the evolutions of synchrotron spectral characteristics, one can then derive the decay index
of the afterglow power-law light-curve as a function of spectral slope and injection-law parameter 
$e$, from where it can be shown that the $e$ value which accommodates the observed decay index 
$\alpha$ of the X-ray hump with its spectral slope $\beta$ is 
\begin{equation}
          e_{(\nu_x<\nu_c)} = \frac{4(3\beta-2\alpha)}{\alpha+3}  \; , \quad 
          e_{(\nu_c<\nu_x)} = \frac{4(3\beta-2\alpha-1)}{\alpha+2}
\label{es0}
\end{equation}
for a homogeneous circumburst medium and
\begin{equation}
          e_{(\nu_x<\nu_c)} = \frac{2(3\beta-2\alpha+1)}{\alpha-\beta} \; , \;
          e_{(\nu_c<\nu_x)} = \frac{2(3\beta-2\alpha-1)}{\alpha-\beta+1}  
\label{es2}
\end{equation}
for a wind, depending on the location of the X-ray domain ($\nu_x$) relative to the cooling 
frequency ($\nu_c$). Equations (\ref{s0}) and (\ref{s2}) for single-$\Gamma_0$ ejecta correspond
to $e \rightarrow \infty$ (i.e. denominators of equations \ref{es0} and \ref{es2} go to zero). 

 The distribution of the injection-law parameter $e$ for 45 X-ray afterglows with humps and for a 
homogeneous medium is shown in Figure \ref{eGs0}. A $\delta\alpha_{x2}=0.2$ and $\delta\beta_{x2}=0.2$ 
uncertainty of the decay index and spectral slope lead to a $\delta e \in (1,2)$ error, thus the
dispersion of indices $e$ shown in Figure \ref{eGs0} is partly intrinsic and partly due to
measurement uncertainties. For a wind medium, about 1/3 of X-ray humps require $e<0$, 
i.e. an accelerating forward shock, in which case equation $t \propto r/\Gamma^2$ for observer time 
may not be valid and $e$ may differ from that given in equation \ref{es2}. For those afterglows 
where $e>0$ (decelerating forward shock), the parameter $e$ has an even wider dispersion than for 
a homogeneous medium.
 
 If the X-ray hump is due to energy injection into a decelerating forward shock, then the ensuing,
faster decay of the X-ray light-curve should be attributed to a transition to a weaker energy
injection (smaller $e$). This defines a Lorentz factor $\Gamma_{break}$ of the ejecta below which
the kinetic energy should be dynamically negligible. From the kinematics of the ejecta--forward-shock
catching-up and the dynamics of the latter, we find that, for a homogeneous medium, 
\begin{equation}
 \Gamma_{break} = 820\, \left( \frac{e+8}{e+2} \right)^{1/2} 
           \left( \frac{E_{53}}{n_0} \right)^{1/8} \left( \frac{t_b}{z+1} \right)^{-3/8} 
\label{gbreak}
\end{equation}
where $E$ is the kinetic energy of ejecta with $\Gamma > \Gamma_{break}$ and $n_0$ is the medium 
particle density in ${\rm cm^{-3}}$. Identifying $E$ with the 10 keV--1 MeV burst output and 
taking $n=1\,{\rm cm^{-3}}$, we obtain the distribution of $\Gamma_{break}$ shown in Figure \ref{eGs0}. 
Given the weak dependence of $\Gamma_{break}$ on $n$, it is unlikely that $\Gamma_{break}$ is universal 
and that its dispersion over 1 decade is due to variations in the circumburst density among bursts.

 The beginning of the X-ray hump is hidden under the GRB tail, hence it is not well-constrained.
Still, the range of the Lorentz factor of the incoming ejecta can be assessed by assuming that energy
injection into the forward shock starts around the end of the burst. Substituting the burst duration
$t_{90}$ in equation (\ref{eGs0}), the ejecta Lorentz factor at the end of the burst is found to be 
$\overline\Gamma_i(t_{90}) = 210 \pm 100$, and the spread in the ejecta Lorentz factor is
$<\Gamma_i(t_{90})/\Gamma_i(t_b)> = 3.6 \pm 1.5$. Thus, to explain the X-ray hump, the ejecta Lorentz 
factor (after internal shocks have ended) must vary by a factor 2--5 to yield an energy injection
into the outflow leading edge that produces the X-ray light-curve hump (consistent with the findings 
of Granot \& Kumar 2006, who used a smaller sample). 
The ratio of the total ejecta kinetic energy to that existing in the forward shock at the end of
the burst is $[\Gamma_i(t_b)/\Gamma_i(t_{90})]^e$, which we find to be between 1.3 and 400, with most
ratios ranging from 2 and 75.

 For about half of the bursts with an X-ray hump, the required power-law evolution of the forward-shock
energy (i.e. the index $e$ shown in Figure \ref{eGs0}) is consistent with that resulting from absorption 
of the dipole electromagnetic emission of a newly-born millisecond pulsar (e.g. Dai \& Lu 1998).
At early times, when the gravitational and dipole radiation losses are less than the pulsar's spin
energy, its rotation frequency $\omega$ is constant and so is the dipole luminosity $L_d \propto \omega^4$. 
Later, when radiation losses amount to a substantial fraction of the pulsar initial spin energy, 
$\omega \propto t^{-1/2}$ and $L_d \propto t^{-2}$. Therefore, only the injection of energy through 
dipole radiation during the $\omega = const$ phase alters the forward-shock dynamics.
The duration of this phase (Zhang \&  M\'esz\'aros 2001) is compatible with that of the X-ray hump. 
From the corresponding evolution of the blast-wave energy ($E \propto t$), the shock adiabaticity 
assumption ($E \propto \Gamma^2 n r^3$), and the radius--time relation for a relativistic source 
($r \propto \Gamma^2 t$), it follows that the absorption of the pulsar dipole radiation leads to
$E \propto \Gamma^{-4}$ (for a homogeneous medium). This injection law is consistent with that inferred
from equation (\ref{es0}) for about half of bursts; still, the injection indices shown in Figure 
\ref{eGs0} are not compatible with a universal value, as could be expected for a millisecond pulsar 
(see also Zhang et al 2006). Assuming that a universal existed, it should be the weighted average 
of the indices of Figure \ref{eGs0}, calculated by taking into account the uncertainty of individual 
values: $\overline{e} = 1.61 \pm 0.10$ for $\nu_c < \nu_x$ and and $\overline{e} = 2.34 \pm 0.07$ for 
$\nu_x < \nu_c$. Therefore, if $e$ were universal, the value inferred from observations of X-ray humps 
is not compatible with that expected for energy injection from a pulsar.

\subsubsection{GRB efficiency}

 The ejecta which produce the X-ray hump emission and those yielding the burst may or may not be 
the same. If the ejecta releasing the burst are only the leading edge of the outflow then, for a
spherical GRB outflow (i.e. a wide jet), the condition that the afterglow flux is not dominated 
by the emission from the burst ejecta (so that we see the X-ray hump) indicates that the afterglow 
ejecta carry more kinetic energy per solid angle than the burst ejecta. This implies 
that the efficiency of the GRB mechanism, defined as the ratio of the GRB output to the kinetic 
energy of the ejecta producing the burst, is larger than that obtained by comparing the afterglow 
kinetic energy with the GRB output and may be, for some bursts, implausibly large or even unphysical 
(above unity). 

  However, this issue of a high burst efficiency would not exist if the GRB outflow were a narrow 
jet whose angular boundary is visible after the burst because, in this case, the kinetic energy 
per solid angle of the burst ejecta can be (much) larger than that of the outflow ejecta (i.e. 
the GRB efficiency can be much less than unity) without the emission from the burst ejecta 
overshining that of the afterglow ejecta after the burst phase (this occurs if the kinetic energy 
of the narrow GRB jet is less than that of the afterglow outflow).

 Thus, for spherical outflows and a plausible efficiency of the GRB mechanism, the burst and 
afterglow outflows should be the same. We investigate now the implications of the burst emission 
arising from ejecta moving at the Lorentz factors shown in Figures \ref{G0s2} and \ref{eGs0}.

\begin{figure*}
\vspace*{-5mm}
\centerline{ \parbox[h]{6cm}{\psfig{figure=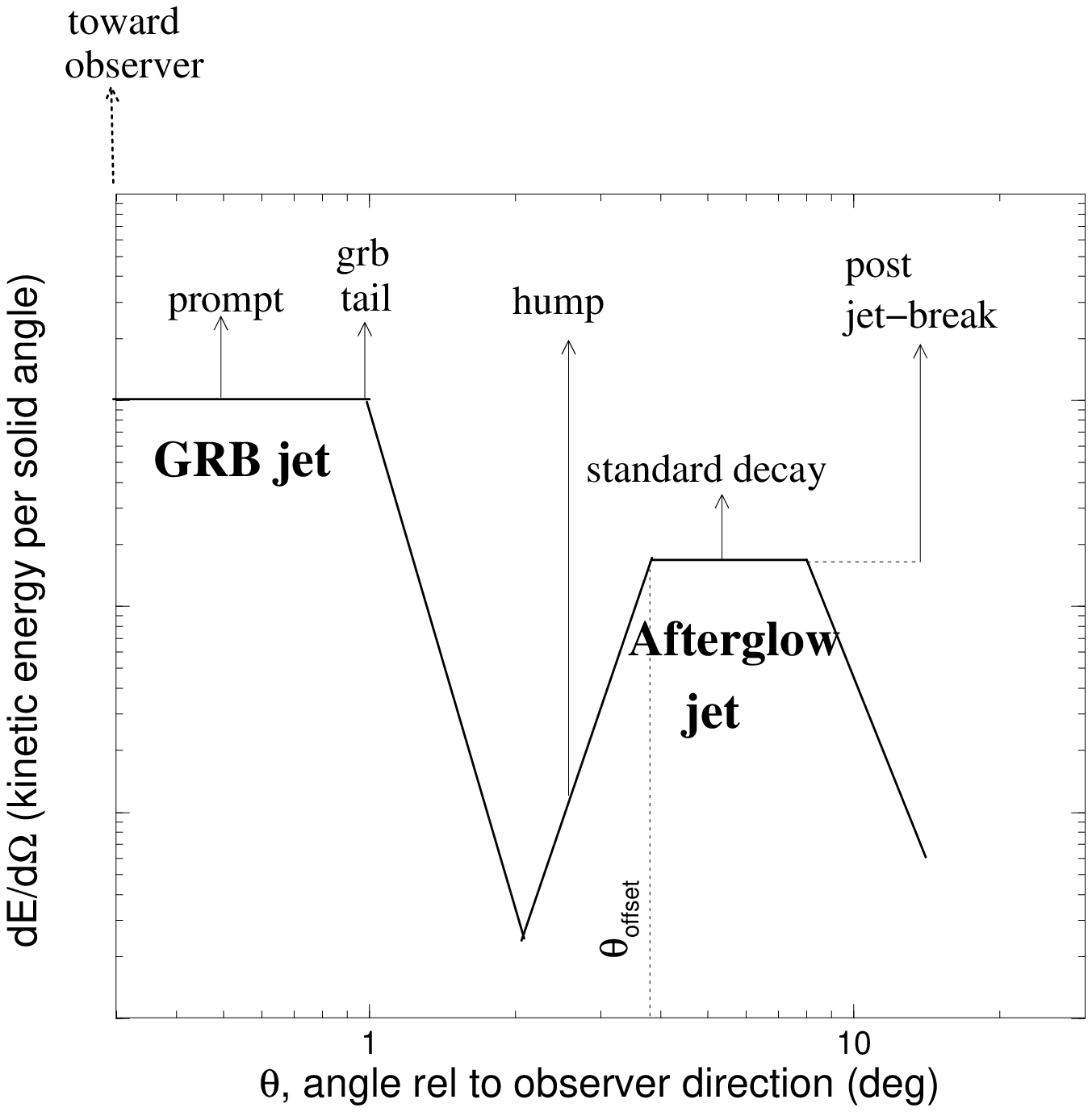,width=5.5cm,height=6.5cm}}
    \hspace*{5mm} \parbox[h]{9cm}{\psfig{figure=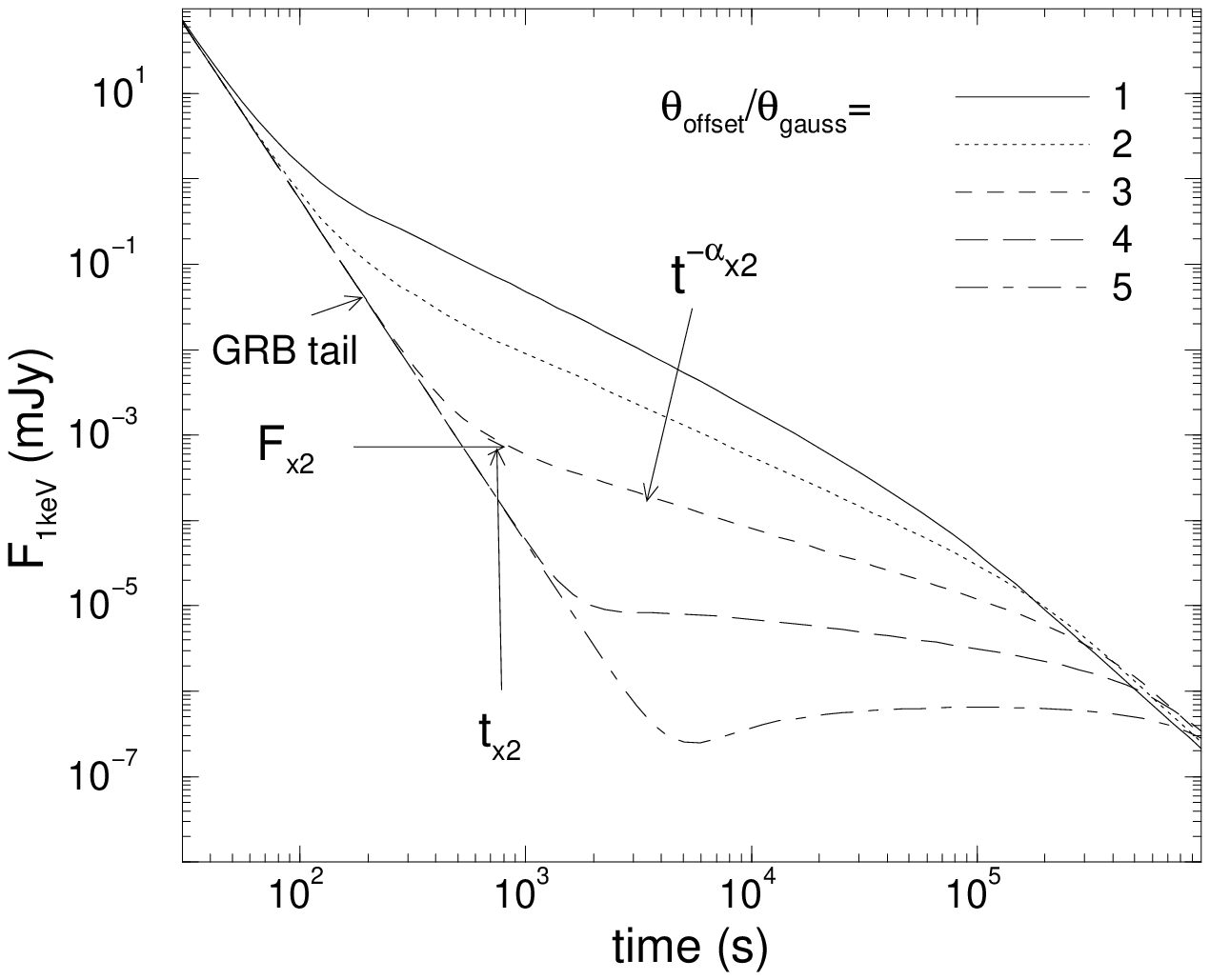,width=7.5cm,height=5.8cm}} }
\caption{ {\it Left panel}: angular distribution of the ejecta kinetic energy per solid angle 
     for a "dual outflow" consisting of a jet moving toward the observer, producing the GRB 
     prompt and tail emissions, and a jet moving at an angle $\theta_{offset}$ off the direction 
     toward the observer, whose emission becomes visible later, yielding the X-ray light-curve hump. 
     {\it Right panel}: as $\theta_{offset}$ increases, the hump emerges at a later time ($t_{x2}$), 
     at a lower flux ($L_{x2}$), and exhibits a slower decay ($t^{-\alpha_{x2}}$), thus a 
     $L_{x2}-\alpha_{x2}$ correlation and a $t_{x2}-\alpha_{x2}$ anticorrelation are expected 
     in this model. Light-curves are for an afterglow jet with a Gaussian angular distribution 
     of the kinetic energy per solid angle. } 
\label{offjet}
\end{figure*}

 The burst ejecta are optically thin to electron scattering if the burst is produced at radii 
larger than
\begin{equation}
 r_e \sim 10^{15}\, (E_{\gamma,53}/\varepsilon_{-1})^{1/2} (\Gamma_0/30)^{-1/2}\; {\rm cm}, 
\label{re}
\end{equation}
where $E_\gamma$ is the GRB output and $\varepsilon$ is the GRB efficiency (normalized to 10 percent). 
For this emission radius, photons from the $\theta < \Gamma_0^{-1}$ visible region arrive at the 
observer over a time 
\begin{equation}
 \delta t = (z+1) r \theta^2/(2\,c) \sim 70\,(E_{\gamma,53}/ \varepsilon_{-1})^{1/2} 
  (\Gamma_0/30)^{-5/2} \; {\rm s}
\end{equation}
for $z=2.5$. Then, to explain the variability 
timescale of Swift bursts, which is $\eta = 10-100$ times smaller than the above $\delta t$, 
the burst emission should be confined to regions of angular extent smaller by a factor $\eta$ 
than the $\Gamma_0^{-1}$ visible region (the "patchy shell" model of Kumar \& Piran 2000). 
To obtain a burst with a high variability, the number of such emitting regions cannot be much larger 
than $\eta$, hence the emitting patches cover a fraction of order $\eta^{-1}$ (Sari \& Piran 1997). 
This means that the condition for optical thinness to electron scattering sets an upper limit of 
1--10 percent for the GRB efficiency, a value consistent with those determined by Granot et al (2006) 
and Zhang et al (2007) for Swift bursts from the burst and afterglow energetics. 

 Pair-formation may alter the GRB spectrum if this spectrum extends above the threshold photon-energy 
$\epsilon_{thr} = \Gamma_0 m_e c^2/(z+1) \sim 4\,(\Gamma_0/30)$ MeV. Assuming that the GRB spectrum
is a power-law ($F_\epsilon \propto \epsilon^{-\beta}$ with $1 < \beta < 2$) up to energies well 
above $\epsilon_{thr}$, it can be shown that a photon of energy $\epsilon$ escapes if the GRB emission 
radius is larger than $r_\pm (\epsilon) = r_\pm (\epsilon_{thr}) (\epsilon/\epsilon_{thr})^{\beta/2}$
with 
\begin{equation}
 r_\pm (\epsilon_{thr}) \sim 10^{15}\, E_{\gamma,53}^{1/2} (\Gamma_0/30)^{-3/4} 
                           (\epsilon_p/{\rm 100\; keV})^{1/4}\; {\rm cm}
\end{equation}
for $\beta = 1.5$, $\epsilon_p$ being the peak energy of the $\nu F_\nu$ burst spectrum.
 For $\beta=1.5$, the condition that the formed pairs are not optically thick to the burst emission 
(e.g. Lithwick \& Sari 2001) leads to a low limit on the emission radius which is 4.4 times larger 
than $r_\pm (\epsilon_{thr})$.

 Taking into account that $r_\pm (\epsilon_{thr}) \propto (E_\gamma \epsilon_p^{\beta-1}/
\Gamma_0^\beta)^{1/2}$, it follows that, depending on the burst spectrum, the condition of optical 
thinness to pair-formation may set a low limit on the GRB source radius higher than that resulting 
from the condition of optical thinness to scattering by the original electron ejecta (equation \ref{re}).
This may lower too much the GRB efficiency, indicating that the underlying assumption of the burst 
and afterglow ejecta being the same is wrong and that the burst outflow should be a narrow jet of 
a kinetic energy per solid angle larger than that of the afterglow outflow.

\subsection{Offset Outflows}
\label{offaxis}

  The X-ray light-curve hump or slow-decay could also arise from a jet whose opening $\theta_0$ 
is less that the offset $\theta_{offset}^{-1}$ between the jet axis and the center--observer direction
(Eichler \& Granot 2006), as illustrated in the left panel of Figure \ref{offjet}. The emission from 
this afterglow is beamed toward the observer when its Lorentz factor has decreased below 
$\theta_{offset}^{-1}$, so that its emission is beamed into cone which includes the direction toward 
the observer. Evidently, this model also requires the existence of an outflow moving toward the observer, 
which produces the GRB emission. This dual-jet model can explain the apparent high GRB efficiency 
provided that the kinetic energy per solid angle in the GRB jet is larger than in the afterglow outflow. 

\begin{figure*}
\vspace*{-4mm}
\centerline{\psfig{figure=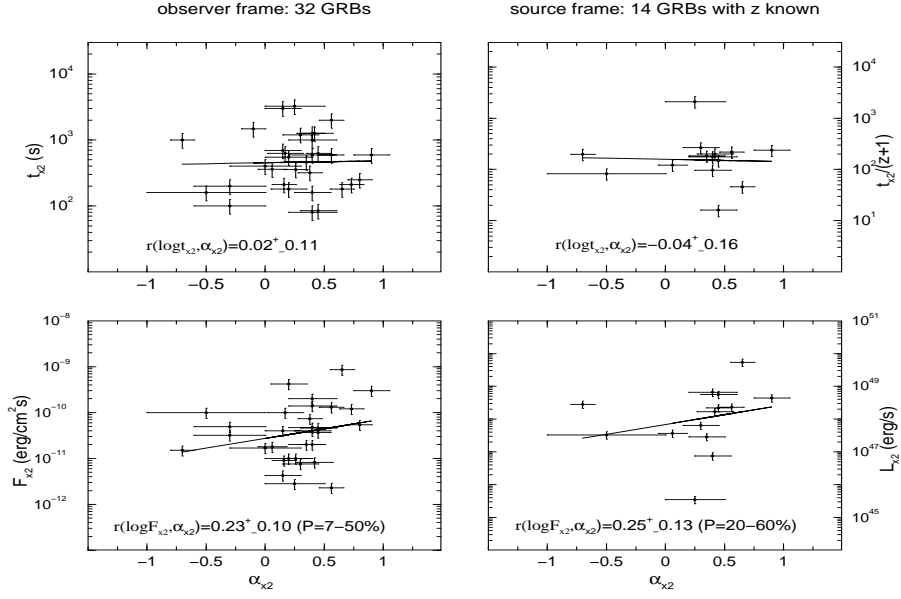,width=12cm,height=8cm}}
\caption{ Test of the correlations among X-ray hump/slow-decay properties expected for the emission 
     from a jet seen from a location outside its opening.
     (or a slow-decay phase). 
   {\it Left panels}: epoch and flux when the afterglow jet emission emerged. 
   {\it Right panels}: corresponding source-frame quantities for those afterglows with known redshift.
   {\it Top panels}: contrary to the expectations for the dual-outflow model, the epoch when the 
     X-ray light-curve hump (or slow decay phase) emerges is not anti-correlated with the decay index. 
   {\it Bottom panels}: as expected in the dual-outflow model, the hump brightness is correlated with 
     the decay index but the statistical significance of this correlation very is low ($P$ is the 
     probability to obtain a linear correlation coefficient $r$ higher than observed in the null 
     hypothesis).
   {\it All panels}: lines indicate linear--log best fits.  } 
\label{aLt}
\end{figure*}

 The emergence of the emission from the afterglow outflow has some specific properties, illustrated 
in the right panel of Figure \ref{offjet}: as the offset angle increases, the light-curve hump should
be seen later, should be dimmer, and should exhibit a slower decay. Therefore, X-ray light-curve
humps arising from jets seen initial off their aperture should exhibit a emergence epoch--decay 
index ($t_{x2}-\alpha_{x2}$) anticorrelation and a brightness--decay index ($F_{x2}-\alpha_{x2}$)
correlation.
 The left panels of Figure \ref{aLt} show that the former anticorrelation is not confirmed with a set 
of 32 bursts whose X-ray hump parameters are well-determined, while the latter correlation may be
true, although it is not manifested at a statistically significant level.
Given that $t_{x2}$ and $F_{x2}$ are dependent on the burst redshift, it is possible that the scatter 
in redshift weakens or completely hides the intrinsic correlations among the luminosity ($L_{x2}$), 
source-frame emergence epoch ($t_{x2}/(z+1))$, and decay index ($\alpha_{x2}$) expected in this model. 
However, as shown in the right panels of Figure \ref{aLt}, restricting the analysis to afterglows with 
known redshift still does not provide observational evidence for the expected correlations.

\section{Chromatic and Achromatic X-ray Humps}

\begin{figure*}
\centerline{\psfig{figure=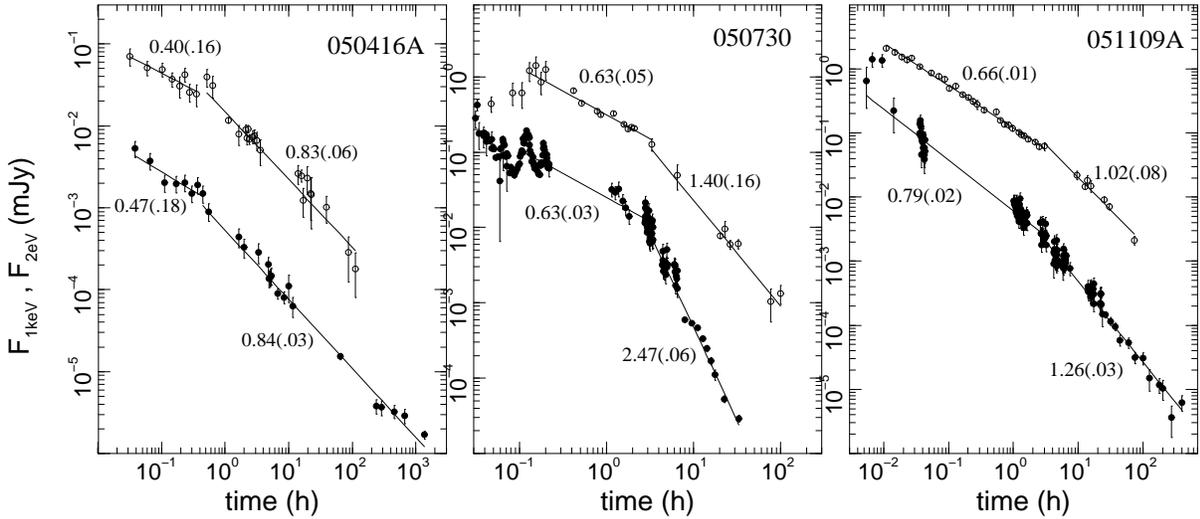,height=7cm,width=16cm}}
\caption{ Optical (top, open symbols) and X-ray (bottom, filled symbols) light-curves for 3 Swift
   afterglows with achromatic X-ray breaks at $\sim 1$ hour. Power-law decay indices and their $1\sigma$
   uncertainties are indicated. The X-ray data are courtesy of P. O'Brien and R. Willingale (Swift). 
   Optical measurements for GRB 050416A are from Holland et al (2007), Soderberg et al (2007);
   for GRB 050730 are from Pandey et al (2006) and GCNs 3717 (A.Blustin), 3778 (S.Kannappan);
   for GRB 051109A are from Yost et al (2007) and GCNs 4230 (F.Huang), 4240 (W.Li), 4259 (K.Misra), 
   4273 (E.Pavlenko), 4259 (K.Kinugasa). 
   Host contributions of $I=22.7\pm 0.1$ (Soderberg et al 2007) and $R=23.70\pm 0.16$ (Yost et al 2007) 
   have been subtracted from the optical fluxes of GRB afterglows 050416A and 051109A, respectively. }
\label{xohump}
\end{figure*}

 If the X-ray hump originates from a long-lived episode of energy injection into the forward shock
then the change in the dynamics of this shock at the end of energy injection should yield a hump at 
other wavelengths as well, i.e. the X-ray light-curve break should be {\sl achromatic}. As shown
in figure 1 of Panaitescu et al (2006b), this expectation is not confirmed by GRB afterglows 050401 
and 050802, whose power-law optical decays do not steepen at the epoch of the X-ray break (around 10 ks) 
but continue unchanged for another 1--2 decades in time. The same figure shows that the optical 
light-curve of GRB 050922C breaks at 1 h while the X-ray data suggest an earlier break epoch. 
GRB afterglows 050319, 050607, and 050713A are also potential cases of {\sl chromatic} X-ray breaks 
occurring at 1--10 h, although the post-break optical coverage is too limited to draw a strong
conclusion.

 Figure \ref{xohump} shows three GRB afterglows with simultaneous optical and X-ray breaks.
For all three, the pre-break optical and X-ray decay indices are comparable, indicating that the 
cooling frequency $\nu_c$ is not between optical and X-ray. The post-break decay indices are equal 
for GRB 050416A, differ by $0.24 \pm 0.09$ for GRB 051109A (which indicates a homogeneous ambient 
medium and $\nu_c$ being between optical and X-ray),
% the latter implies that $\nu_c$ has fallen below the X-ray around the break epoch, i.e. a softening
% of the X-ray spectrum by 1/2 which is larger than observed: $\beta_{x3} - \beta_{x2} = 0.1 \pm 0.1$)
and by $1.1 \pm 0.2$ for GRB 050730. Thus, cessation of energy injection may be compatible with 
the achromatic breaks seen in GRB afterglows 050416 and 051109A, but cannot account alone for the 
discrepancy between the optical and X-ray post-break decay indices of GRB 050730. 
 That $\alpha_{x3} - \beta_{x3} = 1.89 \pm 0.06$ for this burst indicates that its post-break X-ray 
decay is at the limit of the steepest decay allowed for a spherical outflow, that of the large-angle 
emission (released prior to the break), implying that, after the break, the forward shock does not
accelerate electrons to sufficiently high energies to radiate in the X-ray (although consistent with 
being a jet-break, this interpretation is not compatible with the significantly slower post-break 
decay seen in the optical).

 Therefore, there is observational evidence for both chromatic and achromatic breaks at the end of 
the X-ray hump. As chromatic breaks cannot arise from the 
cessation of energy injection in the forward shock, their existence suggests that either this 
scenario is incomplete or that optical and X-ray emission can have, sometimes, different origins. 

 The afterglow optical and X-ray continua are often compatible with a single spectral component 
(after allowing for the cooling frequency to be in between), which suggests a common mechanism for
the optical and X-ray emissions. 
 Then, the decoupling of the X-ray and optical light-curves displayed by the afterglows with chromatic 
X-ray breaks must be attributed to the existence of a spectral break between optical and X-ray. 
% which is such located even before the X-ray break epoch because this light-curve break is not 
% accompanied by a spectral evolution. 
 The spectral break could be the cooling frequency $\nu_c$, however, for the forward-shock emission 
to produce a chromatic X-ray break, the evolution of $\nu_c$ must be non-standard, which in turn 
requires that microphysical parameters evolve. The least contrived scenario is that where the X-ray
hump is due to an episode of energy injection and the evolutions of the micro-parameters is such 
that they "iron out" the optical light-curve break that cessation of energy injection would otherwise 
produce. Interestingly, to accommodate the pre and post-break optical and X-ray decay indices, 
this model requires a wind-like stratification of the circumburst medium (Panaitescu et al 2006b). 

 Genet, Daigne \& Mochkovitch (2007) and Uhm \& Beloborodov (2007) maintain the assumption that the 
optical and X-ray afterglows arise from the same mechanism, which they propose to be the reverse 
shock crossing the incoming ejecta, and attribute the decoupling of the optical and X-ray light-curves 
to the non-standard (i.e. non power-law) distribution of the ejecta electrons with energy that results 
from the continuous injection of fresh electrons and their cooling. To accommodate the observed X-ray 
fluxes, Genet et al (2007) indicate the acceleration of electrons at the reverse shock must be such 
that only a small fraction (around 1 percent) of electrons acquire a large fraction of the dissipated 
energy. 

 Chromatic X-ray breaks may also indicate that the X-ray and optical emissions arise from different
mechanisms. In this venue, Ghisellini et al (2007) propose that the X-ray hump is emission from 
internal shocks, with the optical being dominated by forward-shock emission. This model is more 
likely to be at work in those X-ray humps that exhibit substantial variability, such as the GRB 
afterglow 050904 (Watson et al 2006, Cusumano et al 2007).

 For the remainder of this article, we return to the forward-shock as the origin for the X-ray and
optical afterglow emissions and assume that its microphysical parameters do not evolve.

\section{Standard Decays}
\label{standard}
 
\begin{figure}
\centerline{\psfig{figure=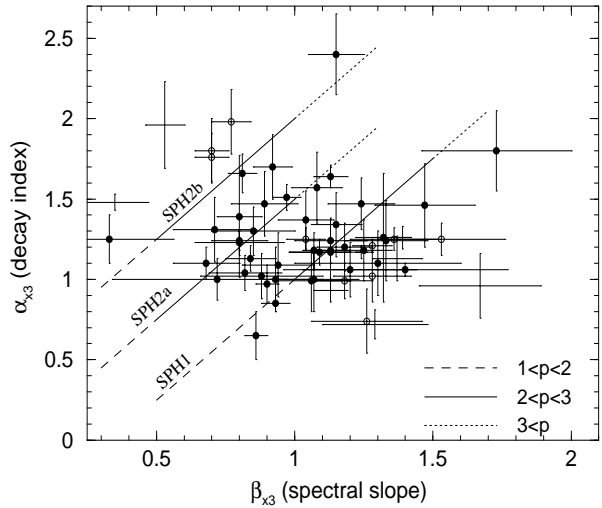,height=7cm,width=8.5cm}}
\caption{ Decay index vs. spectral slope after the X-ray light-curve hump, for 60 Swift afterglows
     (there are more afterglows than shown in Figure \ref{x2} because some X-ray humps were not
      monitored well enough to allow the determination of $\alpha_{x2}$; such afterglows were not
      used for Figure \ref{x2}) .
    Within $2\sigma$, 87 percent of decays are consistent with the expectations of the standard 
     blast-wave model (SPH). However, the decay indices and spectral slopes are not correlated, 
     as might be expected for this model. } 
\label{x3}
\end{figure}

 If we allow for any location of the cooling frequency relative to the X-ray and any of the two possible
stratifications of the circumburst medium, the SPH model (outflow of constant kinetic energy per solid 
angle, whose Lorentz factor is still sufficiently large that the outflow boundary is not yet visible 
to the observer, and with constant microphysical parameters) is consistent at $1\sigma$ with 70 percent
of the standard decays and with 85 percent at $2\sigma$ (Figure \ref{x3}). The SPH1 model (cooling 
frequency below X-ray) is consistent at $2\sigma$ with 60 percent of the observed decays therefore,
for these afterglows, the stratification of the circumburst medium is not constrained.
 
 Despite the correlation between $\alpha$ and $\beta$ expected in any variant of 
the SPH model, the observed decay indices and spectral slope do not display such a correlation: 
$r(\alpha_{x3},\beta_{x3})=-0.18\pm0.10$. If the post-hump X-ray emission arises indeed from 
the forward shock, then this lack of a correlation must be attributed to the scatter in the decay
index for same spectral slope caused by X-ray being either below or above the cooling frequency 
and, perhaps, by the circumburst medium having both possible radial structures.

 The decay index expected for the SPH model is consistent at $2\sigma$ with 75 percent 
of the long-lived slow-decay afterglows, which are shown in Figure \ref{x2} with dotted 
error bars. Still, that more than half of these afterglows lie below the slowest decay 
obtained with the SPH model indicates that the mechanism which reduces the decay of their X-ray 
emission (perhaps the energy injection into the forward shock discussed in the previous section) 
operates until the last measurement. The average decay of the slow-decay afterglows 
($\overline\alpha_{x2} = 0.86$) is faster than that of the afterglows with a hump during the hump
($\overline\alpha_{x2} = 0.37$, Figure \ref{x2}) and slower than after the hump ($\overline\alpha_{x2} 
= 1.24 \pm 0.29$ for the afterglows of Figure \ref{x3}). Thus the difference between the slow-decay
and hump afterglows is that, for the former type, the mechanism that mitigates the decay of the
X-ray emission lasts longer and has a weaker effect than for the latter class.

\section{Jet-Breaks}
\label{jets}

\begin{figure}
\vspace*{-5mm}
\centerline{\psfig{figure=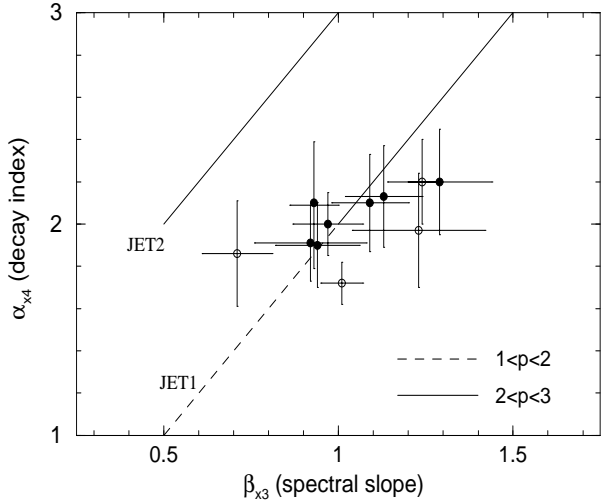,height=7cm,width=8cm}}
\caption{ Decay index vs. spectral slope for Swift afterglows whose X-ray light-curves exhibit 
    a second steepening (or a single steepening, but followed by a steep decay) that can be 
    interpreted as a jet-break. }
\label{x4}
\end{figure}

 The decay of the light-curves of at least half of the well-monitored pre-Swift optical afterglows 
exhibited a steepening at 1 day, which can be interpreted as jet-break. 
In a sample of 78 X-ray afterglows observed by Swift until July 2006, we find 30 whose standard decay 
phase extends up to at least a few days, 19 of them lasting for more than 10 d, while several afterglows 
display a slow $t^{-1.1\pm0.2}$ fall-off until after 30 d. 
The X-ray light-curves of 8 afterglows (GRBs 050315, 050505, 050525A, 050726, 050814, 051221A, 060428A, 
060526) show a second steepening break at 0.1--3 day to a faster decay, which can be interpreted as a 
jet-break (i.e. a light-curve steepening arising from the jet edge becoming visible to the observer). 
For 3 other afterglows (GRBs 050318, 050820A, and 060124) only one break is observed, followed by a 
decay sufficiently steep to warrant a jet-break interpretation as well, yielding a total of 11 X-ray 
afterglows with potential jet-breaks. Willingale et al (2007) identify several other afterglows which 
exhibit a steep decay after the hump. Therefore, the fraction of Swift X-ray afterglows with potential 
jet-breaks is larger than $11/(11+30) \sim 27$ percent of the total number of afterglows monitored for 
more than a few days.

 The jet model (with cooling below X-ray) is consistent at $1\sigma$ with 7 of the 11 post-break decays
(and with all of them within $2\sigma$) listed above, as illustrated in Figure \ref{x4}. For all these 
afterglows, the pre jet-break decay is consistent with the SPH1 model expectations, thus the {\it standard 
jet model} seems able to explain the X-ray light-curve breaks at 0.2--4 d followed by a steep decay. 
Further testing of this model requires optical observations. If a collimated outflow is indeed the reason 
for the late X-ray breaks, then $(i)$ the optical light-curve should exhibit a break at the same time 
(jet-breaks are achromatic) and $(ii)$ the optical and X-ray pre-break ($\alpha_{o3}$, $\alpha_{x3}$) and 
post-break ($\alpha_{o4}$, $\alpha_{x4}$) decay indices should not differ by more than 1/4. This difference 
is expected if the cooling frequency is between the optical and X-ray and could persist after the break 
if the jet does not expand sideways; if the jet spreads laterally, the post-break decay indices should be 
equal even when the cooling frequency is in between optical and X-ray (Sari, Piran \& Halpern 1999). 

\begin{figure*}
\centerline{\psfig{figure=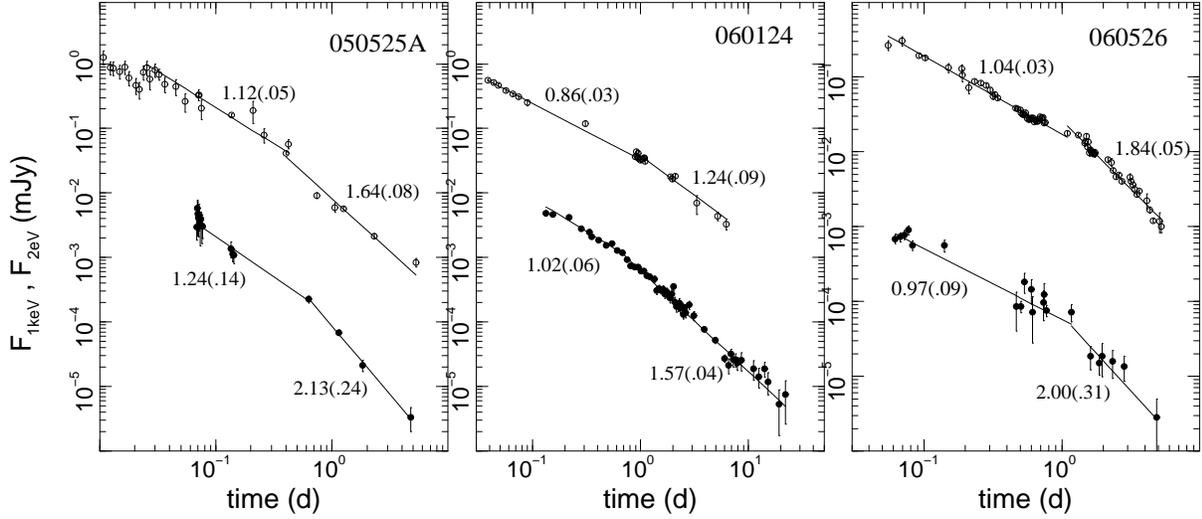,height=7cm,width=16cm}}
\caption{ Optical (top, open symbols) and X-ray (bottom, filled symbols) light-curves for 3 Swift 
   afterglows with potential jet-breaks at $\sim$ 1 day. Power-law decay indices and their $1\sigma$
   uncertainties are indicated. The X-ray data are courtesy of P. O'Brien and R. Willingale (Swift)
   (last two X-ray measurements for GRB 050525A are from Blustin et al 2006).
   Optical measurements for GRB 050525A are from Klotz et al (2005) and Della Valle et al (2006),
   for GRB 060124 from Curran et al (2007) and Misra et al (2007), for GRB 060526 from Dai et al (2007) 
   and GCNs 5166 (E.Rykoff), 5167 (S.Covino), 5169 (C.Lin), 5171 (G.Grego), 5173/5177/5183/5186/5189/5193 
   (I.Khamitov), 5175 (N.Morgan), 5181/5306 (V.Rumyantsev), 5182 (D.Kann), 5185 (K.Baliyan), 5192 
   (F.Terra). }
\label{xojet}
\end{figure*}
  
 For the above set of X-ray afterglows with potential jet-breaks, a sufficiently good optical coverage 
before and after the X-ray break epoch exists only for GRBs 050525A, 060124, and 060526. Their X-ray and 
optical light-curves are shown in Figure \ref{xojet}. As can be seen, the breaks appear achromatic and the 
condition $|\alpha_{x3} - \alpha_{o3}| < 1/4$ is satisfied. 
For GRB afterglow 050525A, $\alpha_{x4}-\alpha_{o4} = 0.49 \pm 0.25$ is above 0.25 at $1\sigma$ and
the break is only marginally consistent with jet origin.
\footnotetext{ There is no evidence for a host galaxy contribution to the post-break optical emission of
 GRB afterglow 050525A and its associated supernova does not dominate the afterglow flux until after 3 d 
  (Della Valle et al 2006)}
For GRB afterglow 060124, $\alpha_{x4}-\alpha_{o4} = 0.33 \pm 0.10$ implies that the cooling frequency
is between optical and X-ray (which is marginally consistent with $\alpha_{x3} = \alpha_{o3} + 0.25$) 
and that the jet does not expand sideways.

\begin{figure}
\centerline{\psfig{figure=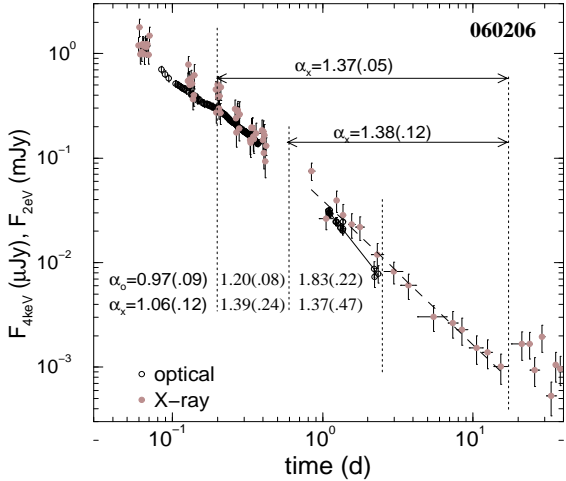,width=7.5cm,height=6.5cm}}
\caption{ GRB afterglow 060206, whose optical light-curve (black dots, solid line fit) suggests 
   the existence of a break at 0.6 d that is not confirmed by the longer post-break X-ray coverage 
   (gray dots, dashed line fit). Power-law decay indices are shown for each fit interval.
   Within their errors, the optical and X-ray decay indices at 1--2.5 d are compatible.
   An achromatic break of $\delta \alpha \simeq 0.25$ is displayed by both light-curves at 0.2 d.
   X-ray data are from Evans et al (2007), optical measurements are from Stanek et al (2007). }
\label{0206}
\end{figure}

 Figure \ref{xojet} shows that achromatic light-curve breaks, as expected for jets, do exist. Burrows 
\& Racusin (2007) show that the X-ray light-curve of GRB afterglow 060206 does not exhibit a steeper 
decay after the epoch (0.6 d) of the optical break identified by Stanek et al (2007). A chromatic
optical break is incompatible with a jet origin and casts doubt over such an interpretation for the
10--15 optical breaks identified for BeppoSAX afterglows. Figure \ref{0206} shows that the post-break 
optical coverage for GRB afterglow 060206 is rather limited (last measurement is at 2.3 d) and that, 
within uncertainty of the X-ray measurements, the optical and X-ray decays at 0.1--2.5 d (when there 
are simultaneous measurements in both bands) are compatible. We suggest that the optical emission 
of this afterglow exhibits a fluctuation at 2 d and that there is no compelling evidence for a 
chromatic optical break at $\sim 0.6$ d. 
                                                                                                               
\section{Conclusions}

 As shown in \S\ref{tail}, less than half of GRB tails are consistent with the delayed, {\it large-angle} 
emission released during the burst. Decays steeper than expected for this model may be due to the burst 
surface-brightness decreasing away from the direction to the observer on an angular scale of $\Gamma^{-1}$. 
The slower decays may be attributed to either the continuation of the burst emission or to the emergence 
of the forward-shock emission before the end of the burst. The former scenario finds support in the X-ray 
flares frequently occurring during the GRB tail, whose short timescale rules out a forward-shock origin 
(e.g. Zhang et al 2006, Burrows et al 2007). The latter scenario is proven to be at work occasionally 
by the long-lived, slow decays which start at the end of the burst and last for days. 

 Three quarters of GRB tails can be explained with the forward-shock synchrotron emission from a very 
narrow jet undergoing lateral spreading (Figure \ref{x1}, right panel), the forward-shock inverse-Compton 
emission from a wide outflow (Figure \ref{x1ic}, left panel), or the inverse-Compton emission from a jet
that does not spread sideways (Figure \ref{x1ic}, right panel). For the jet edge to be visible to the
observer at the end of the burst, the jet half-opening angle must be less than $1^{\rm o}$, which is 2--10 
times smaller than inferred for BeppoSAX afterglows from their $\sim 1$ day optical light-curve breaks 
(e.g. Frail et al 2001, Panaitescu \& Kumar 2001). Together with the interpretation of the X-ray afterglow 
mechanism discussed below, this leads to a {\it double-outflow} forward-shock model (two jets moving in 
the same direction) where the burst emission arises from a narrow, leading jet and the afterglow emission 
from a less relativistic, wider outflow. 

 The X-ray slow-decay phase or hump following the GRB tail can be attributed to the emergence of the 
emission from a newly shocked outflow. This outflow may move along the same direction as that releasing 
the burst (i.e. there is a radial distribution of the ejecta energy), adding energy into the leading 
forward shock (e.g. Nousek et al 2006, Panaitescu et al 2006a, Zhang et al 2006), or slightly off the 
direction toward the observer (i.e. the ejecta energy has an angular distribution), its emission becoming 
visible after it has decelerated enough (Eichler \& Granot 2006). 
 The current sample of afterglows with well-determined X-ray emergence epoch, luminosity, and decay 
index (14 bursts with redshift, 32 in total) do not confirm the correlations expected in the latter 
model (\S\ref{offaxis}, Figure \ref{aLt}), which indicates that energy injection in the forward shock
(\S\ref{enginj}) is the mechanism producing the X-ray hump. 

 For the energy injection model, the duration of the hump requires a 2--5 spread in the Lorentz factor 
$\Gamma$ of the incoming ejecta (see also Granot \& Kumar 2006), with the lowest $\Gamma$ (that of the 
ejecta arriving at the end of the hump) being between 20 and 100 (Figure \ref{eGs0}). A smaller spread 
in the ejecta Lorentz factor is also possible; in the extreme case of a unique $\Gamma$, the slow 
decay observed during the X-ray hump requires a circumburst medium with a wind-like stratification 
(as expected for a massive GRB progenitor) and ejecta Lorentz factors between 10 and 25 (Figure \ref{G0s2}).
As shown in Figure \ref{eGs0}, the increase of the forward-shock energy that accommodates the X-ray hump 
does not obey a universal law, as could be expected if the energy injection were due to absorption of 
the dipole radiation from a newly born millisecond pulsar (Zhang et al 2006).

 Granot et al (2006) and Zhang et al (2007) have found that the GRB output is between 1 and 10 percent
of the forward-shock kinetic energy after the end of the hump (i.e. after energy injection has completed). 
If the forward-shock kinetic energy at the end of the burst is comparable to that of the GRB ejecta, then
the increase by a factor 2--75 of the shock energy during the X-ray hump would imply that the efficiency
of the GRB mechanism is, sometimes, not much below 100 percent. The first caveat here is the assumption
that the ejecta injected during the X-ray hump did not contribute to the GRB emission. The opposite
is possible if the gradual injection of energy in the forward shock is just a kinematic "illusion" 
resulting from the dispersion in the ejecta Lorentz factor at the end of the burst phase. Evidently,
in this situation, the GRB and its tail emissions cannot be from the forward shock, because the ejecta
carrying most of the energy develop this shock only later, during the X-ray hump. Instead, the GRB tail 
must be identified with the large-angle burst emission (whatever is the burst mechanism) or with continuing 
internal shocks.

 The second caveat of the above burst efficiency estimation is that it compares isotropic-equivalent
energies over regions of different apertures, thus the true efficiency of the GRB mechanism may be 
smaller if the ejecta kinetic energy per solid angle over the visible $\Gamma^{-1}$ opening is larger 
during the burst phase than during the afterglow. This could be the case for the double-outflow model 
mentioned above, where the GRB tail is the forward-shock emission from a ultrarelativistic, narrow 
jet ($\theta_{grb} < \Gamma_{grb}^{-1}$) of higher kinetic energy per solid angle and the afterglow
X-ray hump emission is that from the forward shock driven by of a less relativistic, wider outflow 
($\theta_{aglow} > \Gamma_{aglow}^{-1}$) of lower kinetic energy per solid angle. 
% (condition $\E_{grb} \theta_{grb}^2 < \E_{aglow} \theta_{aglow}^2$ should be satisfied to ensure 
% that the observed afterglow emission arises mostly from the wider outflow).

 The X-ray light-curves of 11 afterglows display a break at 0.1--15 d that can be attributed to a jet 
(Figure \ref{x4}). The optical emission for 3 of them was monitored before and after the X-ray break 
epoch, providing evidence for an achromatic steepening of the afterglow emission decay, as expected 
for a jet-break. Their pre-break optical and X-ray decay indices are in agreement with the expectations 
for the standard forward-shock model.
Taking into account the lack of conclusive evidence for achromatic breaks in pre-Swift afterglows, 
optical monitoring of future Swift afterglows will be essential in testing the predictions of the 
widely-used jet model.

 The reverse shock and internal shocks have been recently proposed (Genet et al 2007, Ghisellini et al 
2007, Uhm \& Beloborodov 2007) to be the origin of the X-ray and, perhaps, the optical afterglow emissions, 
motivated by the decoupling of the light-curve decays at these frequencies during the X-ray hump.
The continuity of the X-ray light-curve plateau and the subsequent "standard decay" argues in favour 
of a single mechanism for both. It would be very contrived for internal shocks to yield smooth, 
power-law X-ray light-curves lasting days and weeks, thus this mechanism cannot be a prevalent origin
for the X-ray humps. A sustained injection of ejecta in the reverse shock, characterized by a power-law 
distribution of ejecta mass with Lorentz factor, could account for the long-lived X-ray light-curves, 
but the forward-shock model still remains the most natural explanation for the X-ray afterglows emission
because, in this model, the power-law decaying light-curves are a straightforward consequence of
the power-law deceleration (Lorentz factor versus radius) of the blast-wave caused by its interaction 
with the ambient medium.

\section*{Acknowledgments}
 The author thanks Pawan Kumar, Richard Willingale, and the referee for their helpful suggestions and 
comments. This work made use of data supplied by the UK Swift Science Data Center at the University of 
Leicester.

%\newpage

\end{document}